\documentclass[12pt]{iopart}

\usepackage{iopams}
\usepackage{graphicx}

\begin{document}

\title{Exactly Solved Models and Beyond: a special issue in honour of R J Baxter's 75th birthday}

\author{Murray T Batchelor$^{1,2,3}$, Vladimir V Bazhanov$^2$ and Vladimir V Mangazeev$^2$}

\address{$^1$ Centre for Modern Physics, Chongqing University, Chongqing 400044, China}

\address{$^2$ Department of Theoretical Physics, Research School of Physics and Engineering, Australian National University, Canberra ACT 2601, Australia}

\address{$^3$ Mathematical Sciences Institute, Australian National University, Canberra ACT 2601, Australia}

\begin{abstract}
This is an introduction to {Exactly Solved Models and Beyond}, a special issue
collection of articles published in J.~Phys.~A in honour of R J Baxter's 75th birthday.
\end{abstract}

Rodney J Baxter's pioneering contributions to the study of exactly solved models in statistical mechanics, dating back to the early 1970s,
continue to have a profound impact in both mathematics and physics.
His body of work includes both finding remarkable new solutions
of key models and the invention of powerful techniques for calculating their physical properties.
Baxter's concepts of commuting transfer matrices, functional relations and corner transfer matrices
have inspired developments across a broad spectrum of mathematical physics.
The notion of Yang-Baxter integrability originating from lattice models led to profound advances in quantum field theory,
in knot theory and in the development of quantum groups.
Such integrable models have played a central role in the AdS/CFT correspondence in string theory and are also being
realized in experiments in low-dimensional physics.

Accordingly Baxter has received a number of distinctions and awards throughout his career.
These include:
\begin{itemize}
\item Pawsey Medal, Australian Academy of Science, 1975
\item Elected Fellow of the Australian Academy of Science, 1977
\item Boltzmann Medal, International Union of Pure and Applied Physics, 1980
\item Elected Fellow of the Royal Society of London, 1982
\item Lyle Medal, Australian Academy of Science, 1983
\item Dannie Heineman Prize, American Physical Society, 1987
\item Elected Royal Society Research Professor at Cambridge, 1992
\item Harrie Massey Medal, British Institute of Physics, 1994
\item Centenary Medal, Australian Government, 2003
\item Lars Onsager Prize, American Physical Society, 2006
\item Lars Onsager Lecture and Medal, Norwegian University of Science and Technology, 2006
\item Royal Medal, Royal Society of London, 2013
\end{itemize}

Baxter's work has involved finding brilliant solutions to highly non-trivial mathematical problems.
Colleagues who have glanced over Rodney's shoulder while he calculates will attest to the Baxter
wizardry at deriving and manipulating formulae.
When Baxter started his research 50 years ago the classical culture of ``working with formulas" had been
seemingly forgotten by many mathematicians in their pursuit of abstractions.
Baxter is one of the few who stimulated the renaissance of this culture in modern mathematics.
To perpetuate the culture that his work has also inspired at The Australian National University
for well over half a century,
the Rodney Baxter Endowment has been established to provide a Baxter Fellowship
to support a prominent theoretician, a rising-star or established leader, to visit the ANU each year for a period of 3 months.

This special issue is a collection of articles in honour of Baxter's 75th birthday.
Previous publications celebrating Baxter milestones are the surveys marking his 50th \cite{Barber} and 60th \cite{McCoy} birthdays.
The collection of research papers \cite{Baxter2000} was based around the conference Baxter2000 held in Canberra, Australia in August 2000.
The title of the present collection of articles is shared with the conference Exactly Solved Models
and Beyond held in Cairns, Australia in July 2015.\footnote{See http://baxter2015.anu.edu.au}

\begin{figure}[t]
\begin{center}
\includegraphics[width=0.65\columnwidth]{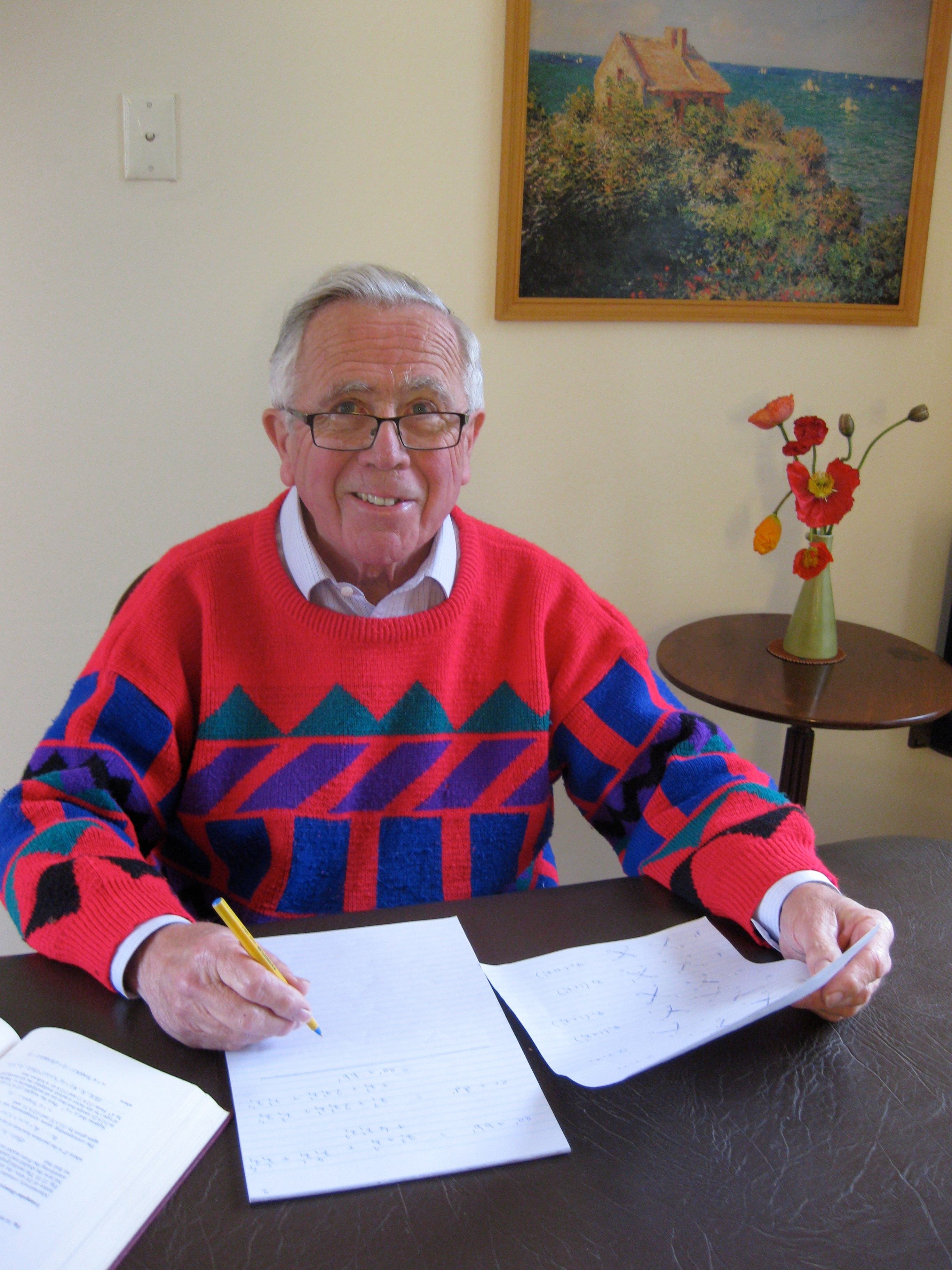}
\caption{Rodney J Baxter taking a break from ``doing sums". Baxter's name is associated with many
mathematical terms, concepts and models in
statistical mechanics and beyond. Photo credit: Elizabeth Baxter.}
\end{center}
\end{figure}

This special issue includes some of Baxter's academic and personal reminiscences
covering three career highlights \cite{Baxter}.
There are also two review articles, one on the early history of the integrable chiral Potts model \cite{Perk},
the other on the impact of Yang-Baxter integrable models in experiments,
from condensed matter to ultracold atoms \cite{BF}.
Moreover, ref.~\cite{BKS} contains a comprehensive review of all known solutions of the star-triangle relation.
Indeed the spread and depth of the articles in this special issue are testament to the broad impact of
Baxter's work and to the vitality
of work in the related areas of mathematical physics.
There are a number of contributions on various aspects of the Yang-Baxter
or star-triangle relation \cite{BKS,YY,Kels,LOZ,Kashaev},
the tetrahedron equation \cite{KOS,KMO} and fusion in the one-dimensional Hubbard \cite{BLN} and RSOS \cite{TP} models.
There are also contributions on CSOS \cite{PA}, SOS \cite{FWZ}, non-unitary RSOS \cite{BR},
non-Abelian anyons \cite{FK},
dimer \cite{MRR}, six-vertex \cite{Martins,TRK}, eight-vertex \cite{NT}, spin-boson \cite{Luk},
generalized Rabi \cite{LB} and dilute orientated loop \cite{VJS} models.

New results are also presented for
scalar products, form factors and correlation functions in integrable models \cite{DGKS,PC,KM,RK,KMNT,JKKS}.
Other topics included are Baxter's $Q$-operators \cite{Q1,Q2},
$Q$-colourings of the triangular lattice \cite{Qcol},
periodic Temperley-Lieb algebras \cite{TL},
the random-cluster model on isoradial graphs \cite{iso},
discrete-time exclusion processes \cite{CMRV},
susceptibility of the square lattice Ising model \cite{GM},
diffusion processes \cite{F},
compressed self-avoiding walks, bridges and polygons \cite{G},
and discrete holomorphicity in the chiral Potts model \cite{holo}.
An exact solution is given for three interacting friendly walks in the bulk \cite{TOR} and
topological defects are considered for the Ising model \cite{Fendley}.

In other contributions,
the Bethe Ansatz method is established for an XXZ type model associated to
quantum toroidal $\mathfrak{g}{{\mathfrak{l}}_{1}}$ \cite{gl1}, the off-diagonal
Bethe Ansatz scheme is discussed for the prototypical XXZ spin \cite{Cao} and
by the same method Bethe states are constructed for the integrable spin-$s$
chain with generic open boundaries \cite{Yang}.
The counting of Bethe roots is also discussed \cite{count1,count2}.

Contributions on the more mathematical side include
a general method to produce flat connections for the two-boundary quantum Knizhnik-Zamolodchikov equations \cite{Isaev},
the study of generalizations of the Rogers-Ramanujan $q$-series \cite{Foda},
quantum B\"acklund transforms and topological quantum field theory \cite{Korff},
dynamics of a $q$-difference Painlev\'e equation \cite{Joshi},
matrix product formula for Macdonald polynomials \cite{Mac},
invariants of the vacuum module associated with the Lie superalgebra ${\mathfrak{gl}}(1| 1)$ \cite{Molev},
and the discussion of diagonals of rational functions occuring
 in lattice statistical mechanics and enumerative combinatorics \cite{diag}.
Closer to the physics side, there are contributions on integrable
pairing models \cite{pair1,pair2}, bosons in a four-well ring \cite{bosons}
and multi-component Fermi gases \cite{Fermi}.

This special collection of articles reflects the profound and ongoing influence of Baxter's work, which we are sure
will continue to inspire further developments in this important and widely influential area of mathematical physics
and beyond for many decades to come.


\section*{References}

\begin{thebibliography}{10}

\bibitem{Barber} Barber M N 1991 Statistical mechanics: A perspective on the work of R J Baxter {\sl Physica A} \textbf{170} 221

\bibitem{McCoy} McCoy B M 2001 The Baxter revolution {\sl J. Stat. Phys.} \textbf{102} 375

\bibitem{Baxter2000} Batchelor M T, Bazhanov V V and Pearce P A (eds) 2001
The Baxter revolution in mathematical physics
{\sl J. Stat. Phys.} \textbf{102} 373-1081


\bibitem{Baxter} Baxter R J 2015 Some academic and personal reminiscences of Rodney James Baxter
{\sl J. Phys. A: Math. Theor.} \textbf{48} 254001

\bibitem{Perk} Perk J H H 2016 The early history of the integrable chiral Potts model and the odd-even problem
{\sl J. Phys. A: Math. Theor.} \textbf{49} 153001

\bibitem{BF} Batchelor M T and Foerster A 2016 Yang-Baxter integrable models in experiments:
from condensed matter to ultracold atoms
{\sl J. Phys. A: Math. Theor.} \textbf{49} 173001

\bibitem{BKS} Bazhanov V V, Kels A P, Sergeev S M 2016
Quasi-classical expansion of the star-triangle relation and integrable systems on quad-graphs
{\sl J. Phys. A: Math. Theor.} \textbf{49}

\bibitem{YY} Yamazaki Y and Yan W 2015
Integrability from 2d ${\mathcal{N}}=(2,2)$ dualities
{\sl J. Phys. A: Math. Theor.} \textbf{48} 394001

\bibitem{Kels} Kels A P 2015
New solutions of the star-triangle relation with discrete and continuous spin variables
{\sl J. Phys. A: Math. Theor.} \textbf{48} 435201

\bibitem{LOZ} Levin A, M Olshanetsky and A Zotov A 2016
Yang-Baxter equations with two Planck constants
{\sl J. Phys. A: Math. Theor.} \textbf{49} 014003

\bibitem{Kashaev} Kashaev R 2016
The Yang-Baxter relation and gauge invariance
{\sl J. Phys. A: Math. Theor.} \textbf{49} 164001

\bibitem{KOS} Kuniba A, Okado M and Sergeev S M 2015
Tetrahedron equation and generalized quantum groups
{\sl J. Phys. A: Math. Theor.} \textbf{48} 304001

\bibitem{KMO} Kuniba A, Maruyama S and Okado M 2016
Multispecies TASEP and the tetrahedron equation
{\sl J. Phys. A: Math. Theor.} \textbf{49} 114001

\bibitem{BLN} Beisert N, de Leeuw M and Nag P 2015
Fusion for the one-dimensional Hubbard model
{\sl J. Phys. A: Math. Theor.} \textbf{48} 324002

\bibitem{TP} Tartaglia E and Pearce P A 2016
Fused RSOS lattice models as higher-level nonunitary minimal cosets
{\sl J. Phys. A: Math. Theor.} \textbf{49} 184002

\bibitem{PA} Perk J H H and Au-Yang H 2016
CSOS models descending from chiral Potts models: degeneracy of the eigenspace and loop algebra
{\sl J. Phys. A: Math. Theor.} \textbf{49} 154003

\bibitem{FWZ} Finch P E, Weston R W and Zinn-Justin P 2016
Theta function solutions of the quantum Knizhnik-Zamolodchikov-Bernard equation for a face model
{\sl J. Phys. A: Math. Theor.} \textbf{49} 064001

\bibitem{BR} Bianchini D and Ravanini F 2016
Entanglement entropy from corner transfer matrix in Forrester-Baxter non-unitary RSOS models
{\sl J. Phys. A: Math. Theor.} \textbf{49} 154005

\bibitem{FK} Frahm H and Karaiskos N 2015
Non-Abelian ${SU}{(3)}_{k}$ anyons: inversion identities for higher rank face models
{\sl J. Phys. A: Math. Theor.} \textbf{48} 484001

\bibitem{MRR} Morin-Duchesne A, Rasmussen J and Ruelle  P 2016
Integrability and conformal data of the dimer model
{\sl J. Phys. A: Math. Theor.} \textbf{49} 174002

\bibitem{Martins} Martins M 2015
The symmetric six-vertex model and the Segre cubic threefold
{\sl J. Phys. A: Math. Theor.} \textbf{48} 334002

\bibitem{TRK} Tavares T S, Ribeiro G A P and Korepin  V E 2015
Influence of boundary conditions on bulk properties of six-vertex model
{\sl J. Phys. A: Math. Theor.} \textbf{48} 454004



\bibitem{NT} Niccoli G and Terras V 2016
The eight-vertex model with quasi-periodic boundary conditions
{\sl J. Phys. A: Math. Theor.} \textbf{49} 044001

\bibitem{Luk} Lukyanov S L 2016
Fidelities in the spin-boson model
{\sl J. Phys. A: Math. Theor.} \textbf{49} 164002

\bibitem{LB} Li Z-M and Batchelor M T 2015
Algebraic equations for the exceptional eigenspectrum of the generalized Rabi model
{\sl J. Phys. A: Math. Theor.} \textbf{48} 454005

\bibitem{VJS} Vernier E, Jacobsen J L and Saleur H  2016
Dilute oriented loop models
{\sl J. Phys. A: Math. Theor.} \textbf{49} 064002





\bibitem{DGKS} Dugave M, G\"ohmann F, Kozlowski K K and Suzuki J 2015
Low-temperature spectrum of correlation lengths of the XXZ chain in the antiferromagnetic massive regime
{\sl J. Phys. A: Math. Theor.} \textbf{48} 334001

\bibitem{PC} Piroli L and Calabrese P 2015
Exact formulas for the form factors of local operators in the LiebÐLiniger model
{\sl J. Phys. A: Math. Theor.} \textbf{48} 454002

\bibitem{KM} Kozlowski K K and Maillet J M 2015
Microscopic approach to a class of 1D quantum critical models
{\sl J. Phys. A: Math. Theor.} \textbf{48} 484004

\bibitem{RK} Ribeiro G A P and Kl\"umper A 2016
Correlation functions of the integrable spin-$s$ chain
{\sl J. Phys. A: Math. Theor.} \textbf{49} 254001

\bibitem{KMNT} Kitanine N, Maillet J M, Niccoli G and Terras V 2016
On determinant representations of scalar products and form factors in the SoV approach: the XXX case
{\sl J. Phys. A: Math. Theor.} \textbf{49} 104002

\bibitem{JKKS} Jiang Y, Komatsu S, Kostov I and Serban D 2016
The hexagon in the mirror: the three-point function in the SoV representation
{\sl J. Phys. A: Math. Theor.} \textbf{49} 174007



\bibitem{Q1} Frassek R 2015
Algebraic Bethe ansatz for $Q$-operators: the Heisenberg spin chain
{\sl J. Phys. A: Math. Theor.} \textbf{48} 294002

\bibitem{Q2} Duval A and Pasquier V 2016
$q$-bosons, Toda lattice, Pieri rules and Baxter $Q$-operator
{\sl J. Phys. A: Math. Theor.} \textbf{49} 154006

\bibitem{Qcol} Vernier E, Jacobsen J L and Salas J 2016
$Q$-colourings of the triangular lattice: exact exponents and conformal field theory
{\sl J. Phys. A: Math. Theor.} \textbf{49} 174004

\bibitem{TL} Jacobsen J L 2015
Critical points of Potts and O(N) models from eigenvalue identities in periodic TemperleyÐLieb algebras
{\sl J. Phys. A: Math. Theor.} \textbf{48} 454003

\bibitem{iso} Beffara V, Duminil-Copin H and Smirnov S 2015
On the critical parameters of the $q \le 4$ random-cluster model on isoradial graphs
{\sl J. Phys. A: Math. Theor.} \textbf{48} 484003

\bibitem{CMRV} Crampe N, Mallick K, Ragoucy E and Vanicat M 2015
Inhomogeneous discrete-time exclusion processes
{\sl J. Phys. A: Math. Theor.} \textbf{48} 484002

\bibitem{GM} Guttmann A J and Maillard J-M 2015
Automata and the susceptibility of the square lattice Ising model modulo powers of primes
{\sl J. Phys. A: Math. Theor.} \textbf{48} 474001

\bibitem{F} Forrester P J 2015
Diffusion processes and the asymptotic bulk gap probability for the real Ginibre ensemble
{\sl J. Phys. A: Math. Theor.} \textbf{48} 324001

\bibitem{G} Beaton N R, Guttmann A J, Jensen I and Lawler G F 2015
Compressed self-avoiding walks, bridges and polygons
{\sl J. Phys. A: Math. Theor.} \textbf{48} 454001


\bibitem{holo} Ikhlef Y and  Weston R 2015
Discrete holomorphicity in the chiral Potts model
{\sl J. Phys. A: Math. Theor.} \textbf{48} 294001

\bibitem{TOR} Tabbara R, Owczarek A L and Rechnitzer A 2016
An exact solution of three interacting friendly walks in the bulk
{\sl J. Phys. A: Math. Theor.} \textbf{49} 154004

\bibitem{Fendley}  Aasen D, Mong R S K and Fendley P 2016
Topological defects on the lattice: I. The Ising model
{\sl J. Phys. A: Math. Theor.} \textbf{49} 354001




\bibitem{gl1} Feigin B, Jimbo M, Miwa T and Mukhin E 2015
Quantum toroidal $\mathfrak{g}{{\mathfrak{l}}_{1}}$ and Bethe ansatz
{\sl J. Phys. A: Math. Theor.} \textbf{48} 244001

\bibitem{Cao} Cao J, Yang W-L, Shi K and Wang Y 2015
On the complete-spectrum characterization of quantum integrable spin chains via inhomogeneous $T-Q$ relation
{\sl J. Phys. A: Math. Theor.} \textbf{48} 444001

\bibitem{Yang} Yang L, Zhang X, Cao J, Yang W-L, Shi K and Wang Y 2015
Bethe states of the integrable spin-s chain with generic open boundaries
{\sl J. Phys. A: Math. Theor.} \textbf{48} 014001

\bibitem{count1} Gainutdinov A M, Hao W, Nepomechie R I and Sommese A J 2015
Counting solutions of the Bethe equations of the quantum group invariant open XXZ chain at roots of unity
{\sl J. Phys. A: Math. Theor.} \textbf{48} 494003

\bibitem{count2} Deguchi T and Giri P R 2016
Exact quantum numbers of collapsed and non-collapsed two-string solutions in the spin-1/2 Heisenberg spin chain
{\sl J. Phys. A: Math. Theor.} \textbf{49} 174001

\bibitem{Isaev} Isaev A P, Kirillov A N and Tarasov V O 2016
Bethe subalgebras in affine Birman-Murakami-Wenzl algebras and flat connections for q-KZ equations
{\sl J. Phys. A: Math. Theor.} \textbf{49} 204002

\bibitem{Foda} Foda A and Welsh T A 2016
Cylindric partitions, ${{\boldsymbol{ \mathcal W }}}_{r}$ characters and the Andrews-Gordon-Bressoud identities
{\sl J. Phys. A: Math. Theor.} \textbf{49} 164004

\bibitem{Korff} Korff C 2016
From quantum B\"acklund transforms to topological quantum field theory
{\sl J. Phys. A: Math. Theor.} \textbf{49} 104001

\bibitem{Joshi} Joshi N and Lobb S B 2016
Singular dynamics of a $q$-difference Painlev\'e equation in its initial-value space
{\sl J. Phys. A: Math. Theor.} \textbf{49} 014002

\bibitem{Mac} Cantini L, de Gier J and Wheeler M 2015
Matrix product formula for Macdonald polynomials
{\sl J. Phys. A: Math. Theor.} \textbf{48} 384001

\bibitem{Molev} Molev A I and Mukhin E E 2015
Invariants of the vacuum module associated with the Lie superalgebra ${\mathfrak{gl}}(1| 1)$
{\sl J. Phys. A: Math. Theor.} \textbf{48} 314001



\bibitem{diag} Bostan A, Boukraa S, Maillard J-M and Weil J-A 2015
Diagonals of rational functions and selected differential Galois groups
{\sl J. Phys. A: Math. Theor.} \textbf{48} 504001




\bibitem{pair1} Links J, Marquette I and Moghaddam A 2015
Exact solution of the $p+{\rm{i}}p$ Hamiltonian revisited: duality relations in the hole-pair picture
{\sl J. Phys. A: Math. Theor.} \textbf{48} 374001

\bibitem{pair2} Lukyanenko I, Isaac P S and Links J 2016
An integrable case of the $p+{\rm{i}}p$ pairing Hamiltonian interacting with its environment
{\sl J. Phys. A: Math. Theor.} \textbf{49} 084001

\bibitem{bosons} Tonel A P, Ymai L H, Foerster A and Links J 2015
Integrable model of bosons in a four-well ring with anisotropic tunneling
{\sl J. Phys. A: Math. Theor.} \textbf{48} 494001

\bibitem{Fermi} Jiang Y, He P and Guan X-W 2016
Universal low-energy physics in 1D strongly repulsive multi-component Fermi gases
{\sl J. Phys. A: Math. Theor.} \textbf{49} 174005




\end{thebibliography}
\end{document}